\documentclass[modern]{aastex61}

\shorttitle{Chemistry of planet formation with OST}
\shortauthors{Pontoppidan et al.}
\begin{document}

\title{The need for a far-infrared cold space telescope to understand the chemistry of planet formation}

\correspondingauthor{Klaus Pontoppidan}
\email{pontoppi@stsci.edu, (410) 338 4744}

\author{Klaus M. Pontoppidan}
\affiliation{Space Telescope Science Institute, }

\author{Edwin A. Bergin}
\affiliation{University of Michigan}

\author{Gary Melnick}
\affiliation{Harvard-Smithsonian Center for Astrophysics}

\author{Matt Bradford}
\affiliation{California Institute of Technology}

\author{Johannes G. Staghun}
\affiliation{Johns Hopkins University}
\affiliation{NASA Goddard Space Flight Center} 

\author{David T. Leisawitz}
\affiliation{NASA Goddard Space Flight Center}

\author{Margaret Meixner}
\affiliation{Space Telescope Science Institute}

\author{Jonathan J. Fortney}
\affiliation{University of California, Santa Cruz}

\author{Colette Salyk}
\affiliation{Vassar College}

\author{Geoffrey A. Blake}
\affiliation{California Institute of Technology}

\author{Ke Zhang}
\affiliation{University of Michigan}

\author{Andrea Banzatti}
\affiliation{Lunar and Planetary Laboratory, University of Arizona}

\author{Tiffany Kataria}
\affiliation{Jet Propulsion Laboratory, California Institute of Technology}

\author{Tiffany Meshkat}
\affiliation{IPAC, California Institute of Technology}

\author{Miguel de Val-Borro}
\affiliation{NASA Goddard Space Flight Center}

\author{Kevin Stevenson}
\affiliation{Space Telescope Science Institute}

\author{Jonathan Fraine}
\affiliation{Space Telescope Science Institute}

%% Note that the \and command from previous versions of AASTeX is now
%% depreciated in this version as it is no longer necessary. AASTeX 
%% automatically takes care of all commas and "and"s between authors names.

%% AASTeX 6.1 has the new \collaboration and \nocollaboration commands to
%% provide the collaboration status of a group of authors. These commands 
%% can be used either before or after the list of corresponding authors. The
%% argument for \collaboration is the collaboration identifier. Authors are
%% encouraged to surround collaboration identifiers with ()s. The 
%% \nocollaboration command takes no argument and exists to indicate that
%% the nearby authors are not part of surrounding collaborations.

%% Mark off the abstract in the ``abstract'' environment. 
\begin{abstract}
At a time when ALMA produces spectacular high resolution images of gas and dust in circumstellar disks, the next observational frontier in our understanding of planet formation and the chemistry of planet-forming material may be found in the mid- to far-infrared wavelength range. A large, actively cooled far-infrared telescope in space will offer enormous spectroscopic sensitivity improvements of 3-4 orders of magnitude, making it possible to uniquely survey certain fundamental properties of planet formation. Specifically, the {\it Origins Space Telescope ({\it OST})}, a NASA flagship concept to be submitted to the 2020 decadal survey, will provide a platform that allows complete surveys of warm and cold water around young stars of all masses and across all evolutionary stages, and to measure their total planet-forming gas mass using the ground-state line of HD. While this white paper is formulated in the context of the NASA Origins Space Telescope concept, it can be applied in general to inform any future space-based, cold far-infrared observatory. 
\end{abstract}

%% Keywords should appear after the \end{abstract} command. 
%% See the online documentation for the full list of available subject
%% keywords and the rules for their use.
%\keywords{protoplanetary disks --- instrumentation}

%% From the front matter, we move on to the body of the paper.
%% Sections are demarcated by \section and \subsection, respectively.
%% Observe the use of the LaTeX \label
%% command after the \subsection to give a symbolic KEY to the
%% subsection for cross-referencing in a \ref command.
%% You can use LaTeX's \ref and \label commands to keep track of
%% cross-references to sections, equations, tables, and figures.
%% That way, if you change the order of any elements, LaTeX will
%% automatically renumber them.

%% We recommend that authors also use the natbib \citep
%% and \citet commands to identify citations.  The citations are
%% tied to the reference list via symbolic KEYs. The KEY corresponds
%% to the KEY in the \bibitem in the reference list below. 

\section{Introduction} \label{sec:intro}
 
Giant planets, planetesimals, as well as comets and Kuiper belt objects form in a rich chemical environment where primordial water and other volatiles mix with new complex chemistry to create a vast diversity of planetary systems. %In analogy with our own solar system, rocky planets and moons of giant planets around other stars probably have a wide range of compositions, in particular of their surfaces and atmospheres. Some may be rich in ices and volatiles, whereas others could be rocky with tenuous atmospheres. 
While the formation and chemical evolution of planets is complex, the initial chemical conditions of the carriers of oxygen, carbon and nitrogen of the planet-forming region at 0.1-10 AU will play a vital role in determining the final makeup of planets \citep{Cleeves14,Bergin15}. Indeed, it is thought that many chemical signatures seen in ancient Solar System material originate in the gas-rich protoplanetary disk phase \citep{Busemann06,Mumma11}.

We do not know whether protoplanetary disks are universally able to seed their planets with water and other volatile species critical for the origin of life, as we know it. In order to measure the abundance and distribution of water in gas and dust actively being incorporated into planetesimals and planets, access to far-infrared lines of water vapor is required. The $5-600\,\mu$m spectra of typical protoplanetary disks are rich in transitions from some of the most abundant solids and molecular gas species, which are not accessible at other wavelengths. 

This white paper is based on a quantitative yield study of a simulated far-infrared spectroscopic survey with the Origins Space Telescope ({\it OST}) of the Orion cluster (Pontoppidan et al. 2018, in prep.). In that work, we argue that a cold, far-infrared telescope in space, with its enormous improvement in sensitivity, will be able to open a new, and previously nearly completely unexplored, window on planet formation and associated chemistry. We identify water and hydrogen deuteride (HD) as some of the most important examples of unique far-infrared tracers of planet formation, but there are many others, including NH$_3$, HDO, hydrides, OI and water ice. HD is a powerful direct tracer of the total planet-forming gas mass, while water is a key tracer of volatiles needed to seed the atmospheres and hydrospheres of potentially habitable planets and satellites.

The connection between water and life is self-evident, and a far-infrared telescope, such as {\it OST}, is a critical need for understanding the origin of habitable worlds. A main connection between planet formation and extra-solar planets is the C/O ratio and its potential link between the composition of the disk at the birth location and the atmosphere of gas giants \citep{Oberg11}. The C/O ratio links the core accretion theory of giant planet formation with the disk chemical composition, which is driven by the condensation/sublimation fronts of major volatile carriers of C (CO, CO$_2$) and O (water). While CO can be accessed from the ground, cold ($T<500$\,K) water can only be reliably detected from space. {\it OST}, with its large spectral grasp, has the capability to constrain the location of the water snowline line hundreds of disks. {\it OST}, through a combination of observations of multiple lines of water vapor and HD, has the capability to observe and constrain the water vapor abundance at, and beyond, the snowline in hundreds of disks, and thereby provide critical missing pieces for measuring C/O ratios as well as absolute abundances (C/H and O/H). This type of information will be transformative as we enter the age of giant planet characterization.

Thus, to understand the full diversity of planet-forming disks and their connection to exoplanetary composition and the availability of water to potential biospheres, it is critical to obtain a sensitive far-infrared spectroscopic census of large samples of protoplanetary disks around stars of all masses, and for a wide variety of environments. We estimate that sensitive far-infrared spectra of 1000 disks are needed, reflecting two dimensions (disk mass and stellar age), 10 bins along each axis, and 10 disks in each bin. Such a survey must be conducted at high spectral resolution ($R=45\,000-200\,000$) in order to detect the lines against the bright infrared continuum, and to spectrally resolve them to measure their spatial distribution by means of fitting Keplerian line profiles. The minimum resolving power is defined as the FWHM line width of the HD 1-0 line at 112\,$\mu$m and the 179.5\,$\mu$m ground-state line of water in a typical protoplanetary disk, viewed at an inclination of $45^{\circ}$ (see Figure \ref{disk_model_overview}). 

\section{Planet-formation and OST}
\label{motivation}

\subsection{Measuring the mass of planet-forming matter in the Galaxy}

The most fundamental quantity that determines whether planets can form, and on what time scale, is the protoplanetary disk mass. Estimates of disk masses are complicated by the fact that the molecular properties of the dominant constituent, molecular hydrogen, lead it to be unemissive at temperatures of 10-30\,K, which characterizes much of the disk mass. To counter this difficulty, the thermal continuum emission of the dust grains \citep{Andrews07} or rotational lines of CO \citep{Williams14} are often used as a proxy for total mass, under the assumption that they can be calibrated to trace the total gas mass. However, sensitive observations have demonstrated that grains have undergone substantial growth, making the determination via dust inherently uncertain \citep{Ricci10}. Further, recent observations indicate that the CO abundance in disks may be orders of magnitude lower than in molecular clouds due to a combination of sequestration below the CO snow line and chemistry \citep{Favre13, Kama16}. These uncertainties are well known, with broad implications on our understanding of the time scale where gas is available to form giant planets, the primary mode of giant planet formation, the dynamical evolution of the seeds of terrestrial worlds, and the resulting chemical composition of pre-planetary embryos.   

The fundamental rotation transition of HD at 112\,$\mu$m has been proposed as a more robust measure of protoplanetary disk mass. HD is a tracer of H$_2$, with an abundance closely matching that of elemental deuterium. The D abundance, in turn, is known from UV absorption spectroscopy of the local interstellar bubble \citep{Linsky98}. Using the Herschel Space Telescope, \citep{Bergin13} measured the gas mass of the TW Hya disk by taking advantage of the fact that the lowest rotational transition of HD is $\sim 10^6$ times more emissive than the lowest transition of H$_2$ for a given gas mass at 20\,K. The subsequent conversion of HD emission to H$_2$ gas mass is well calibrated. Due to Herschel's limited lifetime, the only other deep HD observations obtained were toward six disks, with the result being two additional detections \citep{McClure16}. In Pontoppidan et al. (2018), we find that, with a $5\,\sigma$ line sensitivity of $10^{-21}\,\rm W\,cm^{-2}$, we can measure the minimum-mass solar nebula around solar-type stars out to a distance of Orion (410\,pc) in one hour. Such sensitivity is possible with a 4-5\,K far-infrared telescope, with a collecting area of 25\,$\rm m^{-2}$, and can be achieved by the {\it OST} concept. 

\subsection{Understanding the role of water in planet formation}

\begin{figure*}[ht!]
\centering
\includegraphics[width=12cm]{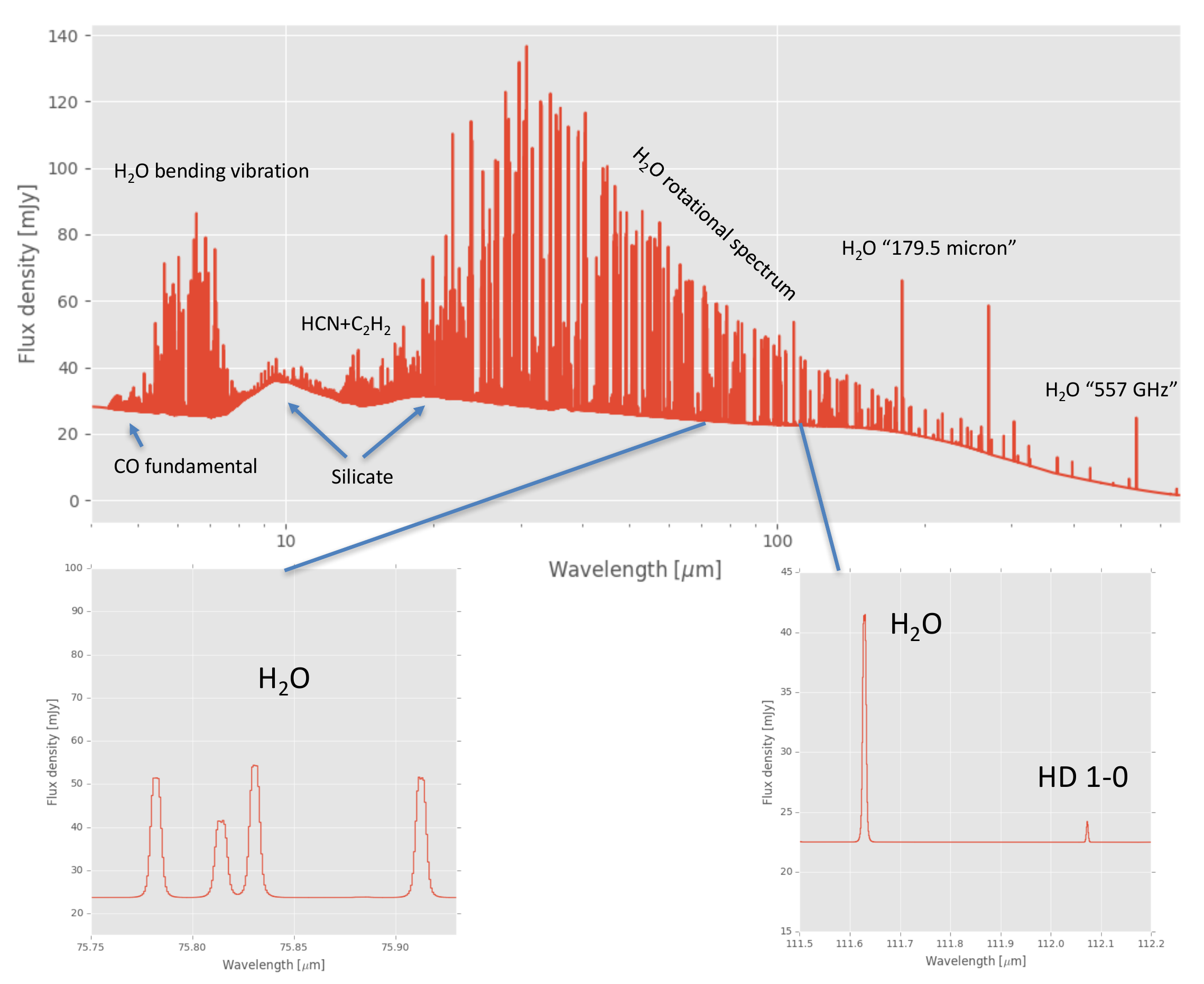}
\caption{A disk model spectrum from Pontoppidan et al. (2018), rendered at 4-660\,$\mu$m, and at a uniform 3\,$\rm km\,s^{-1}$ spectral resolving power. Major spectral features are indicated. The insets show examples of diagnostic molecular lines, including the HD 1-0 line.}
\label{disk_model_overview}
\end{figure*}

To understand the full diversity of planet-forming disks and their connection to exoplanetary composition and the availability of water to potential biospheres, it is critical to obtain an infrared spectroscopic census of the water content in large samples of protoplanetary disks around stars of all masses, and for a wide variety of environments. This can be accomplished by observations of water lines. Based on discoveries by Spitzer, we now know that the mid- to far-infrared spectrum of protoplanetary disks is often dominated by emission from thousands of strong water vapor lines, of which a few hundred are currently detected (see Figure \ref{disk_model_overview}). The corresponding transitions span a very wide range of excitation, or upper level, energies, from less than 100 to 1000s of Kelvin \citep{Carr08,Salyk08,Pontoppidan10b, Banzatti17}. Further, Spitzer observations suggest that there are strong chemical differences in disks around stars of different masses, with water being abundant around solar-mass stars, but much less pronounced around both low-mass and intermediate-mass ($>2\,M_{\odot}$) stars \citep{Pascucci13}. 

This preponderance of water emission lines covering all relevant gas temperatures allows for mapping the distribution of water vapor in protoplanetary disks. Essentially, each line will trace gas in a relatively restricted temperature range, and the combination of many water lines with an appropriate model can map the abundance of water vapor as a function of radius in the disk. Up until now, this has been done for a few disks using unresolved line spectroscopy with Spitzer and Herschel \citep{Zhang13,Blevins16}. At longer wavelengths, Herschel-PACS was able to detect cool water in a few disks at low resolution \citep{Riviere-Marichalar12,Blevins16}. The analysis of such unresolved spectroscopy will always be model-dependent. However, one powerful way to break model degeneracies is to spectrally resolve the lines and use Keplerian motion of the disk to obtain an independent measure of the spatial distribution of the emission for each line. 

\begin{figure}[ht!]
\centering
\includegraphics[width=8cm]{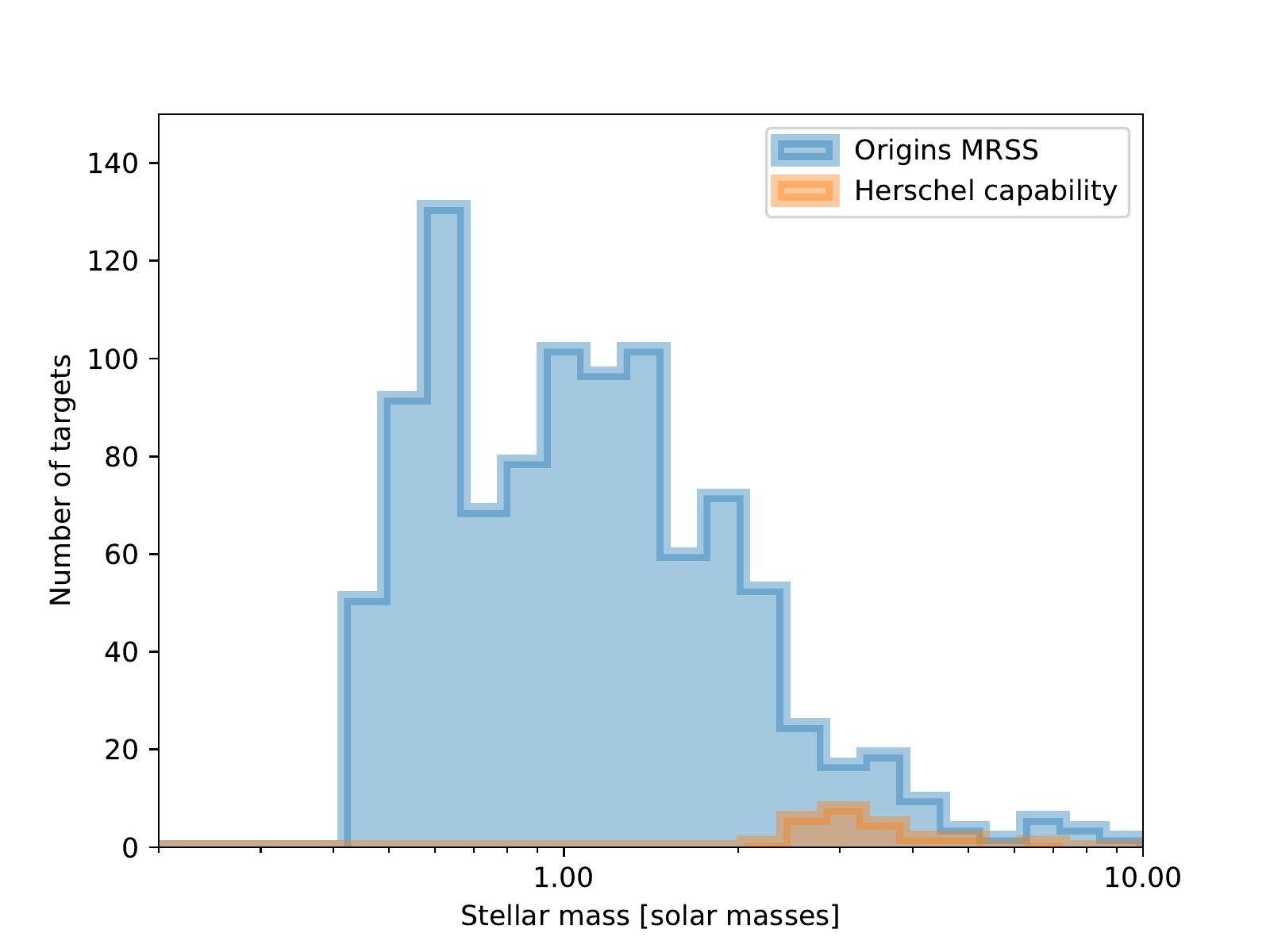}
\caption{Distribution of Orion disks with detectable far-infrared water and HD lines in a notional $\sim$1500 hour program, compared to the distribution of disks that would have been observable with Herschel PACS (at a much lower resolving power of $\sim 2000$) in the same time, demonstrating the improvement possible with {\it OST}.}
\label{orion_dist}
\end{figure}

Indeed, beautifully velocity-resolved Herschel-HIFI observations of the ground-state para and ortho lines at 1113 and 557\,GHz have been used to measure the cold water abundance in the outer, coldest portions of two disks (corresponding to the Sun's Kuiper Belt, 10-30\,K), TW Hya \citep{Hogerheijde11} and DG Tau \citep{Podio13}. At the end of the Herschel mission, we essentially lost the ability to observe water vapor from gas with temperatures in the 50-500\,K range, and therefore our ability to trace the planet- and comet-forming regions at 1-50\,AU, including the location of the snow line. See Figure \ref{orion_dist} for a comparison of Herschel and {\it OST} yields of water and HD. %Such observations are highly complementary to observations of warmer water found nearer to the star in the planet-forming region around the snowline. 

\section{Water and HD in disks in the next two decades}
\label{synergy}

In the 2025-2030 time frame, we expect to have exquisite data sets revealing the physical structure of dust and gas in perhaps $\sim 100$ nearby protoplanetary disks, as observed with ALMA. We will have a census of the amount of water vapor in the surfaces of a similar number of inner disks at 1\,AU from lines tracing hot gas as observed with JWST. JWST will also provide measurements of the composition of a range of exoplanetary atmospheres, and their complement of water and other volatile species. 

Warmer water can be observed from the ground and will be observed by JWST at wavelengths below $\sim 28\,\mu$m, although JWST is unable to observe any water line with upper level energies less than 800\,K, nor will it spectrally resolve the lines. Thus, JWST observations will trace hot water inside of $\sim 1\,$AU and are complementary to {\it OST}. Additionally, a small number of warm-water lines may be observed by ALMA, under very favorable circumstances and for a few bright disks. In the short term, SOFIA-HIRMES will be able to observe water ice via the strong 43\,$\mu$m ice band in many disks, as well as HD 1-0 and warm water in a handful of bright disks. 

%All of these data sets will raise fundamental questions requiring knowledge of the distribution and amount of water beyond 1 AU: How do planets generally acquire their water? Is the water formed in the disk or does it have a primordial origin? Does water play a fundamental role in the ability of disks to form planets? We will not know the answers to these questions without the ability to effectively survey water at all temperatures and phases in planet-forming disks.

\section{Summary and challenges}
A broad, observational census of water colder than 500-1000\,K in large samples of disks requires a space-based, highly sensitive observatory operating at wavelengths of, at least, 30-200\,$\mu$m, with spectral resolving powers of $R \gtrsim 45\,000$ for optimal detection, and $R \gtrsim 200\,000$ to fully resolve the line profiles of cold water and HD 1-0. High spectral resolving power is needed to use emission line profiles to trace the location of the emitting gas in a Keplerian disk. $5\sigma$ line sensitivities of $10^{-21}\,\rm W\,cm^{-2}$ are needed to detect HD in $\sim$1000 disks around stars with masses as low as $0.3\,M_{\odot}$ in one hour (water lines are 5-10 times brighter). This requires a 4-5\,K telescope with a collecting area of $\sim 25\,\rm m^{2}$ (equivalent to that of JWST). A necessary component needed to fully take advantage of the low backgrounds of a cold telescope operating at high spectral resolution are low-noise, multiplexed detectors operating at 30-200\,$\mu$m, and technology for achieving sensitive high-resolution spectroscopy across the far-infrared (30-200\,$\mu$m). 

\bibliographystyle{aasjournal}
\tiny
\bibliography{ms}

%% This command is needed to show the entire author+affilation list when
%% the collaboration and author truncation commands are used.  It has to
%% go at the end of the manuscript.
%\allauthors

%% Include this line if you are using the \added, \replaced, \deleted
%% commands to see a summary list of all changes at the end of the article.
%\listofchanges

\end{document}